\documentclass[doublecol]{epl2}
\usepackage{graphics,graphicx,dcolumn,bm,fleqn,epic,eepic,float}
\usepackage{amssymb,amsmath,multirow,rotate,color,float}
\usepackage[latin1]{inputenc}
\usepackage[dvips]{epsfig}

\graphicspath{{}{images/}}

\def\epsilon{\varepsilon}
\def\theta{\vartheta}

\def\rho{\varrho}

\begin{document}
\title{Anomalous distribution functions in sheared suspensions}

\author{J.~Harting\inst{1} \and H.~J.~Herrmann\inst{2} \and
E.~Ben-Naim\inst{3}}
\shortauthor{J.~Harting \etal}
\institute{
\inst{1} Institute for Computational Physics, Pfaffenwaldring 27, D-70569
Stuttgart, Germany\\
\inst{2} Computational Physics, IfB, Schafmattstr.~6, 
ETH Z\"{u}rich, CH-8093 Z\"{u}rich, Switzerland\\
\inst{3} Theoretical Division and Center for Nonlinear Studies, Los
Alamos National Laboratory, Los Alamos, New Mexico 87545, USA
}

\date{\today}

\abstract{
%\noindent
We investigate velocity probability distribution functions (PDF) of
sheared hard-sphere suspensions. As observed in our Stokes flow
simulations and explained by our single-particle theory, these PDFs can
show pronounced deviations from a Maxwell-Boltzmann distribution.
The PDFs are symmetric around zero velocity and show a
Gaussian core and exponential tails over more than six orders of
magnitude of probability. Following the excellent agreement of our theory
and simulation data, we demonstrate that the distribution functions scale
with the shear rate, the particle volume concentration, as well as the
fluid viscosity.
}

\pacs{02.50.Ey}{Stochastic processes}
\pacs{47.55.Kf}{Particle-laden flows}
\pacs{77.84.Nh}{Liquids, emulsions, and suspensions; liquid crystals}
%\pacs{47.20.Ky}{Nonlinearity, bifurcation, and symmetry breaking}

%\keywords{computer simulations; lattice Boltzmann; non-equilibrium
%distributions; statistical model}

\maketitle

% ============================================================================
\section{Introduction}
\begin{figure}[h]
\centerline{\epsfig{file=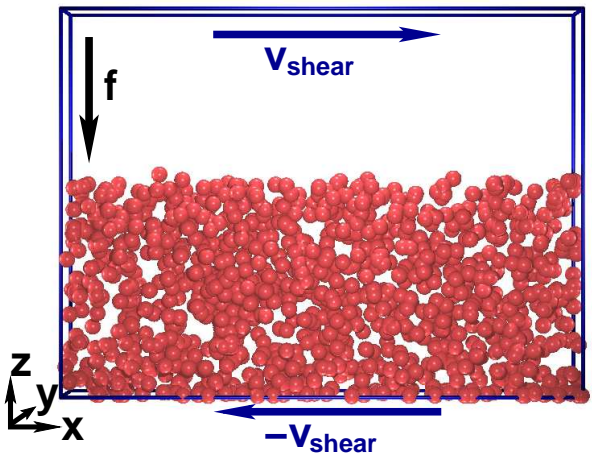,width=0.95\linewidth}}
\caption{\label{Fig:system}(Color online) Sketch of the simulation setup.}
\end{figure}

To describe the statistics of complex systems, often probability
distribution functions (PDF) are utilized.  These distributions have been
found to be of non-Gaussian shape in numerous fields of physics, including
astrophysics~\cite{bib:miesch-scalo:95}, flow in porous
media~\cite{bib:olson-rothman:97}, 
turbulence~\cite{bib:shraiman-siggia:00}, granular
media~\cite{bib:kohlstedt-etal-05,bib:cafiero-luding-herrmann:00,bib:bray-swift-king:07},
or
suspensions~\cite{bib:drazer-koplik-khusid-acrivos-02,bib:abbas-climent-simonin-maxey:06,bib:weeks-crocker-levitt-schofield-weitz:2000}.
However,
the underlying processes are often not
understood. In this letter, we focus on particularly important systems showing
non-Gaussian velocity PDFs, namely sedimenting
hard-sphere suspensions confined between sheared walls (see
Fig.~\ref{Fig:system}). They appear in river beds, blood examinations,
industrial food production, the application of paint, and many more
situations.
Detailed experiments have been performed for more than a hundred years,
but questions about the microstructure or structural
relaxations of the sediment are still not well understood.

Numerous authors have found that the PDF of particle velocities $P(v)$ is
not of similar shape as for an ideal gas, i.e., like a Maxwellian.
Instead, $P(v)$ can show a pronounced non-equilibrium shape, where the
probability of high velocities is substantially
larger~\cite{bib:drazer-koplik-khusid-acrivos-02,bib:abbas-climent-simonin-maxey:06}.
In this letter we present a single-particle theory and simulations to show
that such non-equilibrium distributions can be described as a consequence
of an irreversible driving process, where particles on average gain energy
by one mechanism, but loose energy by another one. Here, energy is
gained from the shear or gravitational forces causing particles to
collide. Contrarily, energy is dissipated due to viscous damping.
This causes $P(v)$ to consist of a Gaussian core and exponential
tails. Even though we focus on a well defined system here, the processes
described are of general nature and can be applicable to ostensibly
different setups.

Experimentally, Rouyer et al.~\cite{bib:rouyer-martin-salin:99} studied
quasi 2D hard-sphere suspensions and found a stretched exponential 
$P(v)$ with concentration dependent exponents between 1 and 2
corresponding to exponential distributions for high concentrations
and Gaussians for small particle counts. These results contradict
theoretical predictions of a transition from exponential to
Gaussian with increasing volume
concentration~\cite{bib:drazer-koplik-khusid-acrivos-02,bib:abbas-climent-simonin-maxey:06}.
%\revision{
However, both experimental and theoretical studies do not present
sufficient statistics over more than 2-4 decades. It is important to note
that if one does not have enough data points for high quality PDFs, final
answers on the nature of the function cannot be given. Indeed, in the
process of analyzing our data we found that even stretched exponentials
can fit PDFs with purely exponential tails and Gaussian centers if only
two to four decades of probability are covered. But as soon as more data
is added, the exponential nature of the tails becomes distinct and it is
impossible to fit the whole PDF with a single function.
%}
 
Here, we overcome such limitations by presenting PDFs
consisting of up to 10$^{10}$ particle displacements each
-- allowing a statistics superior to any previous work. Our data does not
show deviations from purely exponential tails over 6-8 decades of
probability. We show that $P(v_z)$ scales linearly with the shear rate,
volume concentration and viscosity.

The simulated system is a 3 dimensional setup as shown in
Fig.~\ref{Fig:system}. Top and bottom walls are at distance $N_z$ and
sheared with shear rate $\dot\gamma=2v_{\rm shear}/N_z$. All other
boundaries are periodic. A body force $f$ acting on the otherwise
neutrally-buoyant particles can be added to mimic gravity. 
%\revision{
If turned on, this force causes strong density gradients in the
system as can be observed in Fig.~\ref{Fig:system}.
%}
We consider 384
to 1728 initially randomly placed suspended particles of equal radius $a$
corresponding to a particle volume concentration $\phi$ between 6.8\% and
30.7\%. The particle Reynolds number $Re=\dot\gamma 2a/\nu$ is kept
between $0.012$ and $0.07$.

\section{Simulation method}
The simulation method is composed of a lattice Boltzmann solver (LB) for
the fluid and a molecular dynamics (MD) algorithm for the motion
of particles. This approach and recent improvements were
originally introduced by Ladd and
coworkers~\cite{bib:ladd-94,bib:ladd-94b,bib:nguyen-ladd-02,bib:ladd-verberg-01}
and are well established in the
literature~\cite{bib:ladd-94,bib:ladd-94b,bib:nguyen-ladd-02,bib:ladd-verberg-01,bib:nguyen-ladd-04,bib:jens-komnik-herrmann:2004}.
Thus, we only shortly describe it here.

The LB approach allows to calculate long-range hydrodynamic interactions
between particles, by utilizing a discretized version of
Boltzmann's equation~\cite{bib:succi}. Here, positions $x$ are discretized on a 3D lattice
with 19 discrete velocities $c_i$ pointing from a site to its
neighbors. Every $c_i$ is related to a single particle distribution
function $f_i(x)$ which is streamed to neighboring sites at every time
step. After streaming, a collision takes place where the individual
$f_i(x)$ relax towards an equilibrium distribution $f^{\rm eq}_i$.
Local mass and momentum density are given by moments of the $f_i(x)$. 
The movement of suspended particles is modeled by Newton's equation of
motion and appropriate boundary conditions are imposed at solid/fluid
interfaces to exchange momentum. We find a particle radius of $a$=1.25
lattice sites sufficient since for larger radii $P(v_z)$ does not change
significantly anymore, while the computational effort increases
substantially. Also, in low density simulations long-range hydrodynamic
interactions dominate which are correctly reproduced even by small
particles. For dense systems, exact lubrication forces between particle
pairs and between particles and walls are
applied~\cite{bib:ladd-94,bib:ladd-94b,bib:nguyen-ladd-02}. If many
particles come close to each other (less than 0.1 lattice spacings), a
cluster implicit method is used for updating forces in the MD
algorithm~\cite{bib:nguyen-ladd-02}. 
%\revision{
This algorithm does improve the stability of the code and we
carefully checked by comparing simulations with different volume
concentrations and body forces that it has no influence on the shape of
the PDF.
%}

The simulation volume is 64$a$ $\times$ 8$a$ $\times$ 48$a$ and
$\dot\gamma$ is varied between $2.3\cdot 10^{-4}$ and $1\cdot 10^{-3}$ (in
lattice units). The fluid density is kept constant and the kinematic
viscosity is set to $\nu=0.05$ if not specified otherwise.  A single
simulation runs for 6.25 million LB steps, where during the last 5 million
steps the $z$ component of the velocity of every particle is gathered in a
histogram to obtain $P(v_z)$. All distributions are normalized such that
$\int {\rm d}v_z P(v_z)=1$ and $\int {\rm d}v_z v_z^2 P(v_z)=1$ with the
RMS velocity $v_z^{\rm RMS}=\sqrt{\langle v_z^2\rangle}=1$.
%v_x=0.007-0.03
%tau=0.65 shear viscosity: eta=-rho*c_s^2*Dt(1/lambda+0.5)=0.4524
%tau_v=1 fuer bulk visocosity: lambda_v=0 => ??
% particle density =4xfluid density, da massfac=2

\section{Theoretical aspects}
Our theoretical model is based on the balance between viscous
dissipation and shear forcing in steady state. The shear and
the resulting particle collisions are modeled by random, diffusive
forcing. Due to this forcing, the velocity $v_j$ of the $j$th
particle changes as $\frac{dv_j}{dt}=\xi_j$ where $\xi_j$ is a white
noise, $\langle\xi_j\rangle=0$ and
$\langle\xi_i(t)\xi_j(t')\rangle=2D\delta_{ij}\delta(t-t')$.  In
parallel, particles slow down because of the viscous fluid.  In
accordance with the traditional drag law, the velocity decreases
according to $dv/dt=-\beta v$, resulting in the exponential decay
$v(t)=v(0)e^{-\beta t}$ in the absence of forcing.  We model this
viscous damping by reducing each time unit $\Delta t$ the velocity by
a factor $\eta=e^{-\beta}\Delta t$ according to $v\to\eta v$.  In a
sheared fluid, there is a well defined time scale for re-encounters
with the boundary, setting the time scale for the damping
process. This damping process was used by van Zon et al.~as a model
for forced granular
media~\cite{bib:vanzon-kreft-goldman-miracle-swift-swinney:04,bib:larralde:04}.

The velocity distribution obeys the linear but non-local equation
\begin{equation}\label{eq:PDF}
  \frac{\partial P(v)}{\partial t}=
  D\frac{\partial^2P(v)}{\partial^2v}+
  \frac{1}{\eta}P\left(\frac{v}{\eta}\right)-P(v)\mbox{.}
\end{equation}
The first term on the right hand side represents changes due to
diffusive forcing and the next two terms represent changes due to
viscous damping. In our theory, interactions between particles are
represented through the random forcing process reflecting that a
particle undergoes diffusion as influenced by all other particles.  In
steady state, the left hand side of Eq.~\ref{eq:PDF} vanishes. We note
that the shape of $P(v)$ is independent of the diffusion constant
$D$. Indeed, by making the scaling transformation $v\rightarrow
v/\sqrt{D}$ we can eliminate $D$ and assume without loss of generality
that $D=1$. The shape of the distribution depends on the dissipation
parameter $\eta$ alone. The moments
\begin{equation}
%\revision{
M_n=\int dv\,v^nP(v)
%}
\end{equation}
satisfy the
recursion relation
\begin{equation}
%\revision{
M_n={n(n-1)}(1-\eta^n)^{-1}M_{n-2}.
%}
\end{equation}
Since the
distribution is symmetric, $P(v)=P(-v)$, the odd moments vanish and
starting with $M_0=1$, the even moments are
\begin{equation}\label{eq:evenmoments}
  M_{2n}=(2n)!\prod_{k=1}^n\frac{1}{1-\eta^{2k}}\mbox{.}
\end{equation}
Of particular interest is the normalized 4th moment
\begin{equation}
\kappa=M_4/M_2^2={6}/(1+\eta^2).
\end{equation}
The distribution is close to exponential for strong drag,
$\kappa\to6$ as $\eta\to0$ and close to
Gaussian for very weak drag, $\kappa\to3$ as $\eta\to1$. This is
confirmed by the limiting behaviors of all moments 
\begin{equation}
\label{mom-limits}
  M_{2n}\to\left\{
  \begin{array}{l@{\quad\eta\to\,}l}
    (2n)! & 0\mbox{,}\\
    (2n-1)!!(1-\eta)^{-n} & 1\mbox{.}
  \end{array}
  \right.
\end{equation}

The leading large-velocity behavior can be derived by using a heuristic
argument. For sufficiently large velocities, the term
$\eta^{-1}P\left(v\eta^{-1}\right)$ is negligible and hence,
%$\frac{d^2}{dv^2}P(v)=-P(v)$.
\begin{equation}
\frac{d^2}{dv^2}P(v)=P(v).
\end{equation}
Thus, $P(v)$ has an exponential tail, \hbox{$P(v)\sim
\exp\big(-|v|\big)$}. The prefactor can be obtained from the Fourier
transform $F(k)$ that equals an infinite series
\begin{equation}
F(k)=\int_0^\infty dv e^{ikv}P(v)=\prod_{m=0}^\infty[1+k^2\eta^{2m}]^{-1}.
\end{equation}
This expression follows from the steady-state analog of
(\ref{eq:PDF}), \hbox{$(1+k^2)F(k)=F(\eta k)$}.
The simple poles at $\pm i$ closest to the origin imply an exponential
decay, i.e.,% \hbox{$
\begin{equation}
%\revision{
P(v)\simeq{A(\eta)}\exp(-{|v|}),
%}
\end{equation}
%$} 
when
$|v|\to\infty$. Re-summation yields the residue to this pole, and in
turn, the prefactor 
\begin{equation}
%\revision{
A(\eta)=\frac{1}{2}\exp
\left(\sum_{n=1}^\infty\frac{1}{n}\frac{\eta^{2n}}{1-\eta^{2n}}\right).
%}
\end{equation}
The exponential behavior is robust in the limit $v\to\infty$. However,
in the weak drag limit, $\eta\to1$, the exponential behavior holds
only for extremely large velocities. When $\eta\to1$,
($\eta=1-\epsilon, \epsilon\to0$) we expand the denominator in
$A(\eta)$. Keeping only the dominant terms simplifies the sum to
\begin{equation}
%\revision{
\sum_{n=1}^\infty\frac{1}{2\epsilon
n^2}=\frac{\pi^2}{12\epsilon}
%}
\end{equation}
and to leading order, the prefactor is
$A\propto\frac{1}{2}\exp\left[\pi^2/12\epsilon\right]$.  Therefore,
under weak damping, there is a cross-over between a Maxwellian
behavior as follows from (\ref{mom-limits}) and an exponential one,
\begin{equation}\label{eq:Pgaussexp}
  P(v)\sim\left\{
  \begin{array}{@{\,}l@{\quad v\,}c@{\,\epsilon^{-1}}l}
    \exp\left(-\frac{\epsilon v^2}{2}\right) & \ll & \mbox{,}\\
    \exp\left(\frac{\pi^2}{12\epsilon}-|v|\right) & \gg & \mbox{.}
  \end{array}
  \right.
\end{equation}
The two expressions match $P(v)\sim \exp\big(-\epsilon^{-1}\big)$ at
the crossover velocity $v\approx \epsilon^{-1}$. Interestingly, the
crossover to a non-Maxwellian does not affect the leading behavior of
the moments.  

In summary, the theory predicts the non-equilibrium
shape of the PDF as an interplay between energy being injected by a
diffusive thermostat and dissipation due to the fluid drag.  In
general, the high-velocity tail is exponential.  The theoretical
results shown later in this letter are given by a Monte Carlo solution
of the steady state case of Eq.~\ref{eq:PDF}. In these simulations,
$N$ particles are characterized by a velocity $v_i$. The velocities
change through two independent processes: damping and random
forcing. In the damping process, the velocity is reduced by a fixed
factor $v_i\to \eta v_i$. In the forcing process, the particle
velocity changes by a random increment $v_i\to v_i+\xi$ where $\xi$
has zero mean and a unit variance. The steady-state distributions were
obtained using over $10^{10}$ points from simulations with $10^8$
particles.

\section{Results}
\begin{figure}[h]
%\centerline{\epsfig{file=N768fy-072e-4vxvarcollapsenorminset.eps,width=0.98\linewidth}}
\centerline{\epsfig{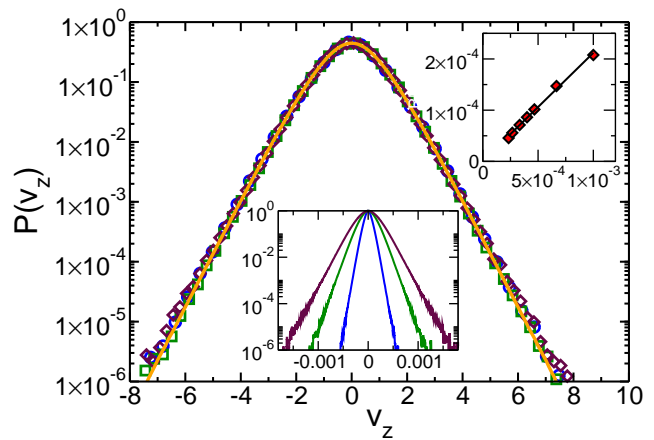}}
\caption{\label{Fig:N768fy-072e-4t0-6vx-var}(Color online) $P(v_z)$ for $f=0.72\cdot 10^{-4}$ corresponding to
a Stokes velocity of $v_s$=$1.4\cdot 10^{-3}$, $\phi=13.6\%$ and shear rates
$3.3\cdot 10^{-4}$, $6.7\cdot 10^{-4}$, and
$1\cdot 10^{-3}$ ($Re$=$0.017$, $0.034$, $0.05$). The solid
line is the steady state solution of Eq.~\ref{eq:PDF} ($\eta=0.73$).
The lower inset shows the unscaled data, where higher $\dot\gamma$ relate to
wider $P(v_z)$. The upper inset shows a linear fit with slope 0.21 of
$v_z^{\rm RMS}(\dot\gamma)$ for the PDFs presented in the main figure
as well as four additional datasets (symbols).}
\end{figure}
First, we consider suspensions with constant $\phi$ and various shear
rates under the influence of a body force $f$. The dependence of $P(v_z)$
on $\dot\gamma$ for three representative values is depicted in
Fig.~\ref{Fig:N768fy-072e-4t0-6vx-var}. 
$P(v_z)$ is symmetric and $\langle
v_z\rangle =0$ for all cases considered in this letter. As shown in the
lower inset, the not normalized $P(v_z)$ widen for higher $\dot\gamma$.
However, a very good scaling is observed: all normalized curves collapse
onto a single one. In the upper inset we show the influence of
$\dot\gamma$ on $v_z^{\rm RMS}$: as expected from the theory,
$\dot\gamma$ only sets a scale for the velocity corresponding to a linear
relation between $\dot\gamma$ and $v_z^{\rm RMS}$. 
$P(v_x)$ (after deducting the shear velocity) and $P(v_y)$ also show
Gaussian cores and exponential tails, although their width and height are
different. Thus, the general shape of distributions in different
directions is essentially identical and they are therefore not shown. 
To obtain an insight into the properties of $P(v_z)$, we compute the
cumulant $\kappa$ and find that for all simulation
parameters studied in this letter it varies between 3.8 and 4.6. Knowing
$\kappa$, we can compute $\eta=\sqrt{6/\kappa-1}$. Due to the large
number of data points in our histograms, we calculate $\kappa$ for periods
of 1 million time steps each and use the arithmetic average of the
last 5 million time steps of a simulation run. We find that $\kappa$
varies by up to 10\% within a single simulation which is of the same order
as the difference of the individual PDFs in
Fig.~\ref{Fig:N768fy-072e-4t0-6vx-var}. Thus, we average the
different curves as well in order to obtain a value for the cumulant to be
utilized for the Monte Carlo solution of the steady state case of
Eq.~\ref{eq:PDF}. For the collapse in
Fig.~\ref{Fig:N768fy-072e-4t0-6vx-var} we get $\eta=0.73$. 
As depicted here, the solid line given by the theory and the
simulation data excellently agree over the full range of six
decades of probability.

\begin{figure}[h]
%\centerline{\epsfig{file=Nviscvarvx0.02fy0-collapsenorm.eps,width=0.98\linewidth}}
\centerline{\epsfig{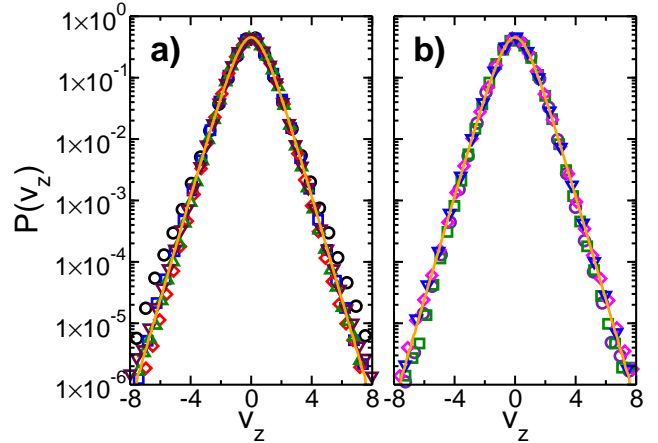}}
\caption{\label{Fig:fy0Nvar}(Color online) $P(v_z)$ for $f=0$,
$\dot\gamma=6.7\cdot10^{-4}$ and $\phi=6.8\%, 13.6\%, 20.5\%, 23.9\%$, and
$27.3\%$, $Re=0.034$ {\bf (a)}. In Fig.~{\bf b)}, $\phi$ is kept at $13.6\%$ and the
kinematic viscosity is set to $\nu=0.017$, $0.05$, and $0.1$
($Re=0.099$, $0.034$, $0.067$).
In both figures, all data sets collapse onto a single
curve and the lines are given by the theory with $\eta=0.69$.
}
\end{figure}

\begin{figure}[h]
%\centerline{\epsfig{file=Vrms-of-phi-visc.eps,width=0.98\linewidth}}
\centerline{\epsfig{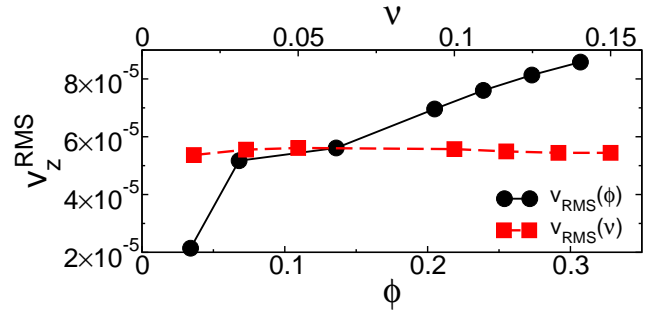}}
%v_RMS(\nu): y = 5.4603e-05 + 1.1974e-06 * x
%v_RMS(\phi): y = 3.3037e-05 + 0.00017565 * x
\caption{\label{Fig:Vrmsphinu}(Color online)
$v_z^{\rm RMS}$ in dependence of $\phi$ (circles) and $\nu$ (squares). 
Data corresponds to $P(v_z)$ as in Fig.~\ref{Fig:fy0Nvar}, but covers a
wider range of $\phi$ and $\nu$. Note the different $x$-axes.}
\end{figure}

Next, we consider
% \revision{
neutrally-buoyant suspended hard spheres
($f=0$)%}
under shear.
The shear rate is kept fixed
% at $\dot\gamma=6.7\cdot 10^{-4}$ 
and the particle concentration is varied between $\phi=6.8\%$ and
$30.7\%$. Due to hydrodynamic interactions, the particles tend to move to
the center of the system, i.e., to an area where the shear is low creating
a depleted region close to the walls. However, this effect does not change the general shape of $P(v)$.
The corresponding normalized $P(v_z)$ are presented in
Fig.~\ref{Fig:fy0Nvar}a.
As depicted in the figure, all PDFs except for the lowest particle
concentration $\phi=6.8\%$ (circles) collapse onto a single curve.  At
very low $\phi$, the tails of $P(v_z)$ are still not fully
converged due to the limited number of particle-particle interactions
taking place within the simulation time frame. Again, the solid line
in Fig.~\ref{Fig:fy0Nvar}a is given by the steady state solution of
Eq.~\ref{eq:PDF} with $\eta=0.69$ being obtained from the 4th moment of
$P(v_z)$. As before, simulation and theory agree very well.
The full circles in Fig.~\ref{Fig:Vrmsphinu} depict the dependence
of $v_z^{\rm RMS}$ on $\phi$. For concentrations of at least
$\phi=13.6\%$,
$v_z^{\rm RMS}(\phi)$ can be fitted by a line with slope $1.8\cdot
10^{-4}$. The disagreement of the linear fit for low $\phi$ is consistent
with the not fully converged PDFs as shown in Fig.~\ref{Fig:fy0Nvar}a.
By keeping all parameters except the kinematic viscosity $\nu$ 
constant, the dependence of $\nu$ on $P(v_z)$ can be studied.
As demonstrated by the squares depicting the dependence of $v_z^{\rm RMS}$
on $\nu$ in Fig.~\ref{Fig:Vrmsphinu}, $v_z^{\rm RMS}$
and thus $P(v_z)$ is independent of the
viscosity. Thus, the steady state curve obtained for different volume
concentrations is identical to the one for different $\nu$ as shown in
Fig.~\ref{Fig:fy0Nvar}b.
It would be interesting to study the influence of the body force 
$f$ on the shape of the PDF. However, 
$f$ and the shear forces are in a subtle interplay since the height of the
steady state sediment depends on both parameters and thus influences the
local concentration. To investigate this
behavior is beyond the scope of this letter.  

\section{Conclusion}
To conclude, the non-equilibrium PDFs reported in this letter are a
consequence of the irreversible nature of the driving process. On average,
particles gain energy by external forces causing particle-particle
collisions but lose energy by viscous damping. Unlike in an ideal gas,
these two mechanisms are not interchangeable. In other words, one cannot
reverse the arrow of time and observe the same behavior. Our theoretical
model captures this irreversibility through the competition between two
non-equivalent driving mechanisms: energy dissipation through a
multiplicative process and energy injection through an ordinary additive
diffusive thermostat.

The theory describes all aspects of the distribution as demonstrated by an
excellent agreement with our coupled LB/MD
simulations of sheared suspensions: the velocity distribution functions
$P(v_z)$ exhibit Gaussian cores and exponential tails over at least 6
orders of magnitude of probability. 
%\revision{
This finding is consistent with experimental results as given for example
in~\cite{bib:kohlstedt-etal-05}.
%}
We also note that the complete shape of the distribution can be
characterized by a single parameter, the normalized 4th moment that has a 
one to one correspondence with the theoretical dissipation parameter.
Further, we confirmed that $P(v_z)$ scales linearly with the particle
volume concentration as well as the shear rate and is independent on the
solvent's viscosity.
 
%\revision{
While various authors report on transitions between Gaussian and
(stretched) exponential
tails~\cite{bib:drazer-koplik-khusid-acrivos-02,bib:abbas-climent-simonin-maxey:06,bib:rouyer-martin-salin:99},
our results do not confirm such transitions. A likely reason for the
previously published results might be the lack of sufficient statistics.
If only two to four orders of magnitude of probability are covered,
$P(v_z)$ can be reasonably well fitted by pure Gaussians or distributions
with (stretched) exponential tails as well. For a conclusive answer on the
shape of the PDF, we found that at least five to six orders of magnitude
of probability and a proper normalization are necessary.  Experiments in
strongly driven granular particles suspended in a fluid provide evidence
of exponential velocity distributions~\cite{bib:kohlstedt-etal-05}.  In
order to close the question of the general validity of the findings
presented in this letter, further high quality experimental or numerical
data would be needed.
%}  

\begin{acknowledgments}
We thank A.J.C.~Ladd for providing his simulation code and for fruitful
discussions. This work was
supported by the DFG priority program ``nano- and microfluidics'', US-DOE grant
DE-AC52-06NA25396, and the ``Landesstiftung Baden-W\"urttemberg''. H.J.~Herrmann
thanks the Max Planck prize. We thank P.L.~Krapivsky for fruitful
discussions and the Institute for Pure and Applied Mathematics at
University of California, Los Angeles for hospitality. The computations
were performed at the J{\"u}lich Supercomputing Centre. 
\end{acknowledgments}

%\bibliographystyle{abbrv-unsrt-notitle}
%\bibliographystyle{eplbib}
%\bibliography{main,jens-pub}

\end{document}